\documentstyle[aps,prl,epsf]{revtex}
\begin{document}
\twocolumn[\hsize\textwidth\columnwidth
\hsize\csname@twocolumnfalse\endcsname
\draft
\title{The effects of nuclear spins on the quantum relaxation of the magnetization\\
for the molecular nanomagnet Fe$_{8}$}

\author{W. Wernsdorfer$^1$, A. Caneschi$^2$, R.~Sessoli$^2$, D. Gatteschi$^2$,
A.~Cornia$^3$, V.~Villar$^4$, C.~Paulsen$^4$, }
\address{
$^1$Lab. L. N\'eel, associ\'e \`a l'UJF, CNRS, BP 166,
38042 Grenoble Cedex 9, France\\
$^2$Department of Chemistry, University of Firenze, via Maragliano 77,
50144 Florence, Italy\\
$^3$Department of Chemistry, University of Modena, 41100 Modena, Italy\\
$^4$CRTBT, associ\'e \`a l'UJF, CNRS, BP 166,
38042 Grenoble Cedex 9, France
}
\date{\today}
\maketitle
\begin{abstract}
The strong influence of nuclear spins on resonant quantum tunneling in the molecular
cluster Fe$_8$ is demonstrated for the first time by comparing the relaxation rate of the
standard Fe$_8$ sample with two isotopic modified samples:
(i) $^{56}$Fe is replaced by $^{57}$Fe, and
(ii) a fraction of $^1$H is replaced by $^2$H.
By using a recently developed "hole
digging" method, we measured an intrinsic broadening 
which is driven by the hyperfine fields. Our measurements are in good agreement 
with numerical hyperfine calculations.
For $T >$ 1.5 K, the influence of nuclear spins on the relaxation rate is
less important, suggesting that spin--phonon coupling dominates the relaxation rate.
% at higher temperature.
\end{abstract}
\bigskip
\pacs{PACS numbers: 75.45.+j, 75.60Ej}
\vskip1pc]
\begin{figure}[b]
\centerline{\epsfxsize=5 cm \epsfbox{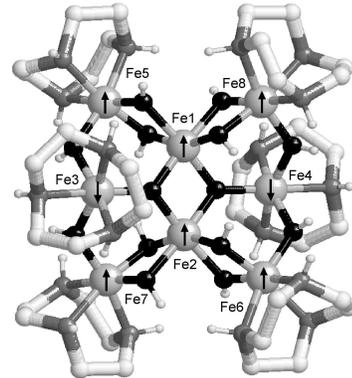}}
\caption{Schematic view of the magnetic core of the Fe$_8$ cluster.
The oxygen atoms are black, the nitrogen gray and carbon atoms are white.
For the sake of clarity only the
hydrogen atoms that are exchanged with deuterium are shown as small
spheres. The
arrows represent the spin structure of the ground state $S=10$
as experimentally determined
through polarized neutron diffraction experiments
%~\cite{Pontillon99}
.}
\label{fig_Fe8}
\end{figure}

\narrowtext
Mesoscopic quantum phenomena are actively investigated both for fundamental science
and for future applications, for instance in quantum computing. Magnetic molecular
clusters are among the most promising candidates to observe mesoscopic quantum
phenomena~\cite{Thiaville99,JMMM_200}. One of the most prominent examples is an
octanuclear
iron(III) cluster, called Fe$_8$ (Fig.~\ref{fig_Fe8}), with a spin ground
state of $S = 10$~\cite{Barra96}. Below 360~mK, the magnetization 
relaxes through a pure tunneling
process giving rise to a stepped hysteresis cycle~\cite{Sangregorio97}.
Furthermore, the tunnel splitting $\Delta$ of Fe$_8$ 
shows periodic oscillations
when a transverse magnetic field is
applied along the hard axis~\cite{Science99}, 
a long searched phenomenon in magnetism
associated with the Berry phase~\cite{Loss92}, and predicted several years
before~\cite{Garg93}. Since $\Delta$ is extremely
small for the ground state tunneling, ca. $10^{-7}$~K at $H = 0$, the
tunneling process should occur only in an extremely narrow magnetic field
range, ca.
$10^{-8}$~T, and should be practically unobservable. However, a recent theory
proposes that the tunneling is mediated by fluctuating hyperfine fields
generated by
magnetic nuclei~\cite{Prok_Stamp98}, but direct experimental evidence is so
far lacking.

In order to study the influence of nuclear spins, we increased the 
hyperfine coupling by the substitution of $^{56}$Fe with $^{57}$Fe, 
and decreased it by the substitution of $^1$H with $^2$H. 
We found that the relaxation rate of magnetization in the 
tunneling regime shows a clear isotope effect which we attribute 
to the changed hyperfine coupling.

The crystals of the standard Fe8 cluster, $^{\rm st}$Fe$_8$ or Fe$_8$,
[Fe$_8$(tacn)$_6$O$_2$(OH)$_{12}$]Br$_8$.9H$_2$O
where tacn = 1,4,7- triazacyclononane, were prepared as reported by
Wieghardt {\it et al.}~\cite{Wieghardt84}. 
For the synthesis of the
$^{57}$Fe-enriched sample, $^{57}$Fe$_8$, 
a 13 mg foil of 95$\%$ inriched $^{57}$Fe 
was dissolved in a few drops of
HCl/HNO$_3$ (3 : 1) and the resulting solution was used as the iron
source in the standard procedure. The $^2$H-enriched Fe$_8$ sample,
$^{\rm D}$Fe$_8$, was crystallized from pyridine-d$_5$ and D$_2$O (99$\%$)
under an inert atmosphere at 5$^{\circ}$C by using a non-deuterated
Fe(tacn)Cl$_3$ precursor. The amount of isotope exchange was not
quantitatively evaluated, but it can be reasonably assumed that the H
atoms of H$_2$O and of the bridging OH groups, as well as a part
of those of the NH groups of the tacn ligands are replaced by deuterium
while the aliphatic hydrogens are essentially not affected. The crystalline
materials were carefully checked by elemental analysis and
single-crystal X-ray diffraction.

The magnetic measurements were made on single-crystal 
samples by using an array of
micro-SQUIDs~\cite{hole_dig_PRL}, which measure the magnetic field
induced by the magnetization of the crystal. The advantage of this
magnetometer lies
mainly in its high sensitivity and fast response, allowing short-time
measurements down to 1 ms. Furthermore the magnetic field 
can be changed rapidly and along any direction.

\begin{figure}[t]
\centerline{\epsfxsize=7.5 cm \epsfbox{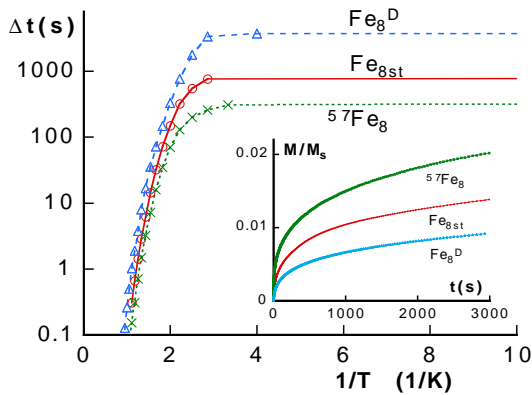}}
\caption{Comparison of the relaxation rates of three different Fe$_8$ samples.
The time $\Delta t$ needed to relax one percent of the
saturation magnetisation $M_{\rm s}$ is plotted versus inverse
temperature $1/T$. The initial magnetisation $M_{\rm init}$ was reached
by a fast cooling in zero
applied field. The relaxation was measured in a field of $\mu_0H_{z}$ = 42~mT. 
In the inset typical relaxation curves ($M$ versus time) recorded for 
the three sample at $T$ = 40 mK are shown.
}
\label{fig_tau}
\end{figure}

\begin{figure}[t]
\centerline{\epsfxsize=7.5 cm \epsfbox{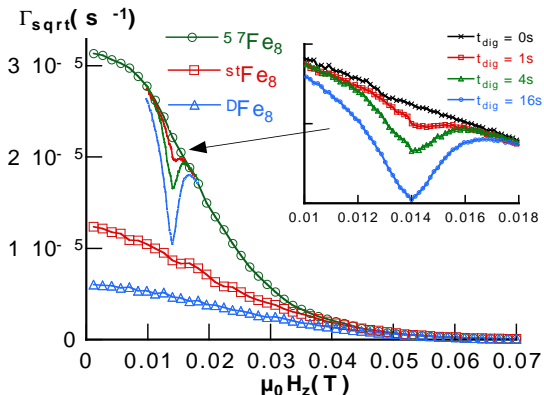}}
\caption{Comparison of the short time relaxation rates of three different Fe$_8$ 
samples at $T$ = 40 mK with $H_{\rm trans} = 0$ and $M_{\rm init} = 0$. 
The inset displays an typical example of a hole which was dug into the distribution
by allowing the sample to relax for the time $t_{\rm dig}$  at $\mu_0H_{\rm dig} =$ 14~mT.
}
\label{gamma_H}
\end{figure}

In order to avoid the influence of the crystal shape~\cite{Ohm98} the
relaxation was
measured starting at an initial magnetization $M_{\rm init} = 0$ where
intermolecular dipolar
interactions lead to a field distribution with a 
width of about 50~mT~\cite{hole_dig_PRL}.
A small field $H_z$ was then applied and the 
relaxation of magnetization was measured
(inset of Fig.~\ref{fig_tau}). 
For all three samples the
relaxation was clearly non-exponential and could be adjusted to a
$\sqrt{\Gamma_{\rm sqrt} t}-$law for the short time regime 
($t < 100$ s) for $T <$ 0.4  K. Fig.~\ref{gamma_H} displays the 
field dependance of $\Gamma_{\rm sqrt}(H_z)$.
The relaxation of the 
three samples at 40~mK are strikingly
different from each other. The $^{57}$Fe$_8$ sample is the fastest relaxing one
whereas the $^{\rm D}$Fe$_8$ shows the slowest relaxation rate.
As a complete theory of the
relaxation behavior
of crystals of molecular clusters is still missing~\cite{remark1},
we plot in Fig.~\ref{fig_tau}
the time needed to relax one percent of the saturation magnetization
$M_{\rm s}$ as a
function of the inverse temperature $1/T$. 
Relaxation and ac susceptibility
measurements at
$T~>$~1.5~K showed no clear difference between 
the three samples suggesting that
above this temperature the relaxation is predominately due to
spin-phonon coupling~\cite{Fort98,Leuenberger99}.
Although the increased mass of the isotopes changes the 
spin--phonon coupling, we believe that this effect is small.

In principle, the change of masse does not change the crystalline field of the 
Fe ions, i.e. the anisotropy constants.
Experimentally, this is confirmed with measurements below $T <$ 0.35~K,
where spin--phonon coupling is negligible, by two observations: 
(i) relative positions of the resonances as a function of the 
longitudinal field $H_z$ are unchanged~\cite{remark2}, and (ii) all three samples
showed the same period of oscillation of $\Delta$ as a function of
the transverse field $H_x$~\cite{Science99}, a period which is 
very sensitive to any change of the anisotropy constants. Finally, 
we point out that the mass is increased in {\it both} isotopically 
modified samples whereas 
the effect on the relaxation rate is opposite.

A deeper insight into the relaxation mechanism can be achieved by using our
recently
developed hole digging method which allows us to estimate the hyperfine level
broadening~\cite{hole_dig_PRL}. Starting from the well defined
magnetization state
$M_{\rm init} = 0$, and after applying a small field $H_{\rm dig}$,
the sample is let to relax for
a time $t_{\rm dig}$, called digging field and digging time, respectively.
During the digging
time, a small fraction $\Delta M_{\rm dig}$ of the molecular spins tunnel
and reverse the
direction of their magnetization. Finally, a field $H$ is applied to
measure the short time
square root relaxation rate $\Gamma_{\rm sqrt}$~\cite{hole_dig_PRL,Ohm98}.
The entire
procedure is then repeated to probe the distribution at other fields 
yielding the field
dependence of the relaxation rate
$\Gamma_{\rm sqrt}(H,H_{\rm dig},t_{\rm dig})$ which is more or less
proportional to the number of spins which are still free for tunneling.
The result of this
procedure is that a very sharp "hole" is dug into the rather broad
distribution of $\Gamma_{\rm sqrt}$~\cite{hole_dig_PRL}. 
A typical example is shown in the inset of Fig.~\ref{gamma_H}.

\begin{figure}
\centerline{\epsfxsize=7.5 cm \epsfbox{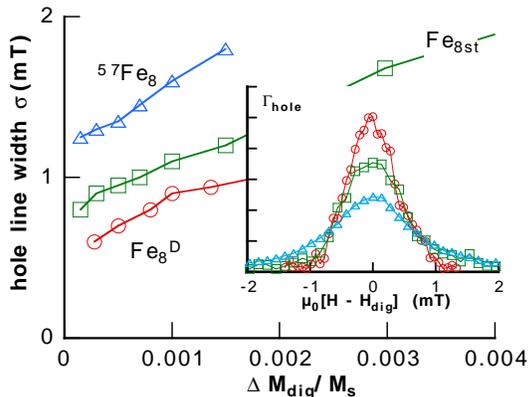}}
\caption{Hole line width $\sigma$ as a function of reversed 
fraction $\Delta M_{\rm dig}$ to dig the hole. The
initial thermal distribution of diploar fields was reached by a fast
cooling in zero applied
field. In order to reduce the line width the 
hole was dug at a longitudinal field of $\mu_0H_{z}$ = 42~mT,  
and in the presence of a transverse field of 
$\mu_0H_{\rm trans}$ =  200~mT applied along  the hard axis. 
In the inset shows typical hole shapes observed 
at $\Delta M_{\rm dig}  = 0.001~M_{\rm s}$
}
\label{fig_dig}
\end{figure}

In the limit of very short digging times, the difference between the
relaxation rate in the
absence and in the presence of digging,
$\Gamma_{\rm hole} =
\Gamma_{\rm sqrt}(H,H_{\rm dig},t_{\rm dig}=0) -
\Gamma_{\rm sqrt}(H,H_{\rm dig},t_{\rm dig})$, is
approximately  proportional to the number of molecules which reversed their
magnetization during
the time $t_{\rm dig}$. $\Gamma_{\rm hole}$ is characterized
by a width that we call the hole line
width $\sigma$. In order to find a hole line width that is close to the
hyperfine level
broadening, all the effects that can broaden the measured hole width must
be reduced. The
experimental condition giving the smallest line width was found for
hole digging in
the tails of the dipolar distribution and for small initial magnetization. 
Under these conditions the spins that tunnel are statistically far from
each other allowing us to measure tunneling in the diluted limit. 
In addition, we applied a
transverse field of $\mu_0H_{\rm trans}$ = 200~mT parallel to the hard axis
which reduces the
tunnel rate allowing us to dig very tiny holes. Fig.~\ref{fig_dig} displays
the hole line width
$\sigma$ as a function of the reversed 
fraction $\Delta M_{\rm dig}/2M_{\rm s}$ of
molecular spins. A linear extrapolation of
$\sigma$ to $\Delta M_{\rm dig}/2M_{\rm s} = 0$ gives
$\sigma_0$ which is directly associated to the hyperfine level
broadening\cite{Prok_Stamp98}. Experimentally we found 
$\sigma_0$ to be $0.6\pm0.1$,
$0.8\pm0.1$, and $1.2\pm0.1$~mT, 
for $^{\rm D}$Fe$_8$, $^{\rm st}$Fe$_8$,
and $^{57}$Fe$_8$, respectively.

The isotope effect, observed here for the first time in magnetic
nanostructures, clearly
point out the role of the magnetic nuclei in the relaxation of the
magnetization. An
evaluation of the hyperfine fields in the three different samples is
therefore necessary.
The hyperfine interaction between the total spin $S$ of the cluster and the
magnetic nuclei
can be decomposed into the sum of terms related to the interactions between
the magnetic
moment $I_i$ of the $i-$th nucleus and the individual spin $S_j$,
assumed to be localized on the $j-$th iron centre~\cite{Bencini90}:
\begin{equation}
        H_{hf} = \sum_{i} {\bf S}á{\bf A}_iá{\bf I}_i
                = {\bf S} \sum_{i}á\left( \sum_{j} c_j {\bf A}_{ij}á\right)
{\bf I}_i
        \label{sum_1}
\end{equation}
The projection coefficients $c_j$ in Eq. (\ref{sum_1}) depend on the
wave function of the ground state $S=10$,
which can be calculated by diagonalizing the $S=10$ block
$(6328 \times 6328)$ of the exchange
spin-Hamiltonian matrix of $H= \sum_{j \not= k}{\bf S}_j{\bf S}_k$.

With the exchange-coupling  parameters that
best reproduce the temperature dependence of the magnetic
susceptibility~\cite{Pontillon99}, the spin
configuration depicted in Fig.~\ref{fig_Fe8} provides a large
contribution (of ca. 70$\%$) of the ground state $S = 10$.
According to this picture, which has been recently confirmed by polarized
neutron diffraction data~\cite{Pontillon99}, the projection coefficients
are $c_3 = c_4 = -5/22$ for the spins
pointing down and $c_1 = c_2 = c_5= ... = c_8 = 8/33$ for the remaining
iron spins. 

The hyperfine interaction described by the ${\bf A}_{ij}$ tensors is
both through-space (dipolar) and through bond (contact) in nature.
The dipolar components can be easily calculated with the
point dipolar approximation. The most important coupling 
with the $^1$H nuclei are those of the bridging 
OH groups having ${\bf A}_{ij}$ constants 
as large as 0.045~mT,
while the hyperfine coupling with N and Br nuclei does 
not exceed 0.005 and 0.003~mT respectively.
An order of magnitude for the contact terms was estimated using 
a Density-Functional Theory 
calculation~\cite{Frisch98} at
the B3LYP level~\cite{Becke93} on a model symmetric
dimer $[($NH$_3)_4$Fe$($OH$)_2$Fe$($NH$_3)_4]^{4+}$ .

The hyperfine interaction generates a field which splits
the $m_S = \pm10$ states. Since each
nuclear spin I splits each state into $2I+1$ sublevels,
the number of sublevels generated by
the coupling of 18 $^{14}$N atoms $(I =1 )$,
8 $^{79,81}$Br atoms $(I = 3/2 )$, and 120 $^1$H atoms (I =1/2)
present in $^{\rm st}$Fe$_8$ is prohibitively large,
being $3^{18} \times 4^8 \times 2^{120}$. We therefore made an
approximation taking into account only the most significant terms. In 
fact the total line width goes as the geometric sum of the individual contributions
and therefore the largest contributions dominate over the smaller ones.
We evaluated the gaussian broadening determined by the 12 $^1$H nuclei
of the OH groups, assuming them
equivalent with a contact hyperfine coupling constant $A_{\rm cont}$ = 0.05~mT, of
the 18 $^1$H nuclei
of the NH groups assuming $A_{\rm cont}$ = 0.025~mT, and of the 14 N nuclei with $A_{\rm cont} = 0.2$~mT.
These values introduced in Eq. (\ref{sum_1}) gave gaussian lines
with widths at half-height of 0.2, 0.15, and 0.4~mT, respectively. 
By combining these we
estimate a resulting gaussian distribution with a line width of 0.5~mT, in
acceptable agreement with the experimental value of 0.8~mT. 

%In our calculation we have in fact 
%considered average values, but as stated 
%above nuclei with a larger hyperfine coupling than 
%the average are dominating in the resulting line width.

\begin{figure}
\centerline{\epsfxsize=8 cm \epsfbox{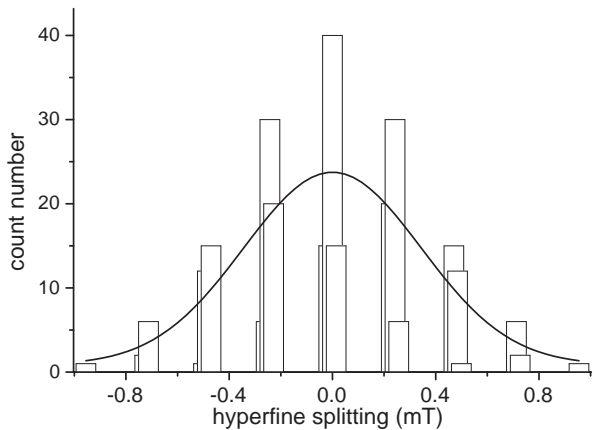}}
\caption{Calculated histogram of level splitting due to the coupling 
with $^{57}$Fe $I = 1/2$ nuclear spins in the Fe$_8$ clusters. 
In equation (\ref{sum_1}) ${\bf A}_{ij}$ constant have been 
assumed to be 1.0~mT for $i = j$ and zero for $i \neq j$. 
The solid line represents the best fit using a gaussian line
}
\label{fig_histo}
\end{figure}

The effect of the $^{57}$Fe nuclear spins in the enriched
samples was estimated by assuming
that each nucleus only feels its own electron spin. 
Therefore Eq. (\ref{sum_1}) is simplified
as only the terms with $i = j$ are different from zero.
Using  $A(^{57}$Fe) = 1.0~mT, in agreement
with reported data~\cite{McGarvey66}, we calculate the stick diagram reported in
Fig.~\ref{fig_histo} which arises from the coupling of six 
equivalent $^{57}$Fe $I = 1/2$ with $A = 8/33 \times$1.0~mT and
two other nuclear spins $I = 1/2$ with coupling $A = -5/22 \times$ 1.0~mT. The related histogram can be fitted with a gaussian whose line width is ca 0.8~mT.
If we consider the experimental line width of the resonance,
ca. 0.8~mT, of $^{\rm st}$Fe$_8$ and we add the
contribution of the $^{57}$Fe nuclei we obtain $\Delta H \approx$ 1.1~mT
in close agreement with the experimentally observed value of 1.2~mT.

The partial substitution for the $^1$H nuclei of the OH and NH groups with
the less
magnetic $^2$H isotopes leads to a reduction 
of the line width which in our calculations is
estimated to be ca. 0.1~mT which should be compared with the experimental
narrowing of ca. 0.25~mT. The difference between the calculated and 
observed reduction of the line width is similar to the smaller calculated 
line width compared to the observed one of the $^{\rm st}$Fe$_8$ sample. This may come from an underestimation of the interactions with the $^1$H nuclei. 

The present data show the fundamental role of the 
nuclear spins in the relaxation of the
magnetization of Fe$_8$ in the quantum regime. 
Indeed, this is in contrast to the familiar role of isotope substitution which are 
generally associated with phonon coupling and thus proportional to the mass of the
nuclei. Here we show that it is the magnetic moment of the nuclei 
which is important at temperatures well
above those at which nuclear spin polarisation is observed.

We are indebted to A. Bencini for his help in DFT calculations, 
and to P.C.E. Stamp, I. Tupitsyn, N.V. Prokof'ev, and J. Villain 
for many fruitful and motivating discussions. 
The financial contributions of Italian MURST,
CNR, PFMSTAII, and the French DRET are acknowledged.

\newpage
\widetext

\begin{references}
\bibitem{remark1} The Prokof'ev Stamp theory is only valid in the pure quantum
regime~\cite{Prok_Stamp98}, i.e. below 360~mK for Fe$_8$~\cite{Sangregorio97}.
\bibitem{remark2} We observed a small shift of the resonances of the 
order of magnitude of 1~mT, positive for $^{57}$Fe$_8$ and negative for
$^{\rm D}$Fe$_8$ ($M_{\rm init}  = -M_{\rm s}$). 
This can also be attributed to the modified 
hyperfine fields. However, a quantitative measurement is complicated by the fact that it is impossible to have two crystal with exactly the same shape, i.e. the same internal fields.
\end{references}
\end{document}